\documentclass[conference]{IEEEtran}
\IEEEoverridecommandlockouts
\usepackage{amsmath,amssymb,amsfonts}
\usepackage{algorithmic}
\usepackage{graphicx}
\usepackage{textcomp}
\usepackage{xcolor}
\usepackage{url}
\usepackage[hidelinks]{hyperref}
\usepackage[nolist,nohyperlinks]{acronym}
\usepackage[sort,compress]{cite}
\usepackage{layouts}
\usepackage{booktabs}
\usepackage{etoolbox}
\usepackage{microtype}
\usepackage{multirow, makecell}
\usepackage{svg}
\usepackage{threeparttable}
\usepackage{tikz}
\usepackage{pgfplots}
\usepgfplotslibrary{external}
\usetikzlibrary{fit, backgrounds, positioning, arrows.meta, shapes.misc, shapes.geometric, shapes.symbols, calc}
\ifCLASSOPTIONcompsoc
    \usepackage[caption=false, font=normalsize, labelfont=sf, textfont=sf]{subfig}
\else
\usepackage[caption=false, font=footnotesize]{subfig}
\fi
\makeatletter
\patchcmd{\@makecaption}
  {\scshape}
  {}
  {}
  {}
\makeatother
\begin{acronym}
\acro{gin}[GIN]{Graph Isomorphism Network}
\acro{wl}[WL]{Weisfeiler-Lehman}
\acro{lstm}[LSTM]{Long Short-Term Memory}
\acro{gnn}[GNN]{Graph Neural Network}
\acro{tcn}[TCN]{Temporal Convolutional Network}
\acro{cnn}[CNN]{Convolutional Neural Network}
\acro{rnn}[RNN]{Recurrent Neural Network}
\acro{mpnn}[MPNN]{Message-Passing Neural Network}
\acro{rmse}[RMSE]{Root Mean Square Error}
\acro{mae}[MAE]{Mean Absolute Arror}
\acro{dtw}[DTW]{Dynamic Time Warping}
\acro{enb}[eNB]{Evolved NodeB}
\acro{gru}[GRU]{Gated Recurrent Unit}
\acro{ran}[RAN]{Radio Access Network}
\acro{qos}[QoS]{Quality of Service}
\acro{rh}[RH]{Radio Head}
\acro{es}[ES]{Edge Server}
\acro{rc}[RC]{Regional Cloud}
\acro{embb}[eMBB]{Enhanced Mobile Broadband}
\acro{urllc}[URLLC]{Ultra-Reliable Low Latency Communication}
\acro{mmtc}[mMTC]{Massive Machine Type Communication}
\acro{rl}[RL]{Reinforcement Learning}
\acro{drl}[DRL]{Deep Reinforcement Learning}
\acro{ilp}[ILP]{Integer Linear Programming}
\acro{sla}[SLA]{Service Level Agreement}
\acro{mdp}[MDP]{Markov Decision Process}
\acro{ppo}[PPO]{Proximal Policy Optimization}
\acro{cc}[CC]{Computing Core}
\acro{milp}[MILP]{Mixed-Integer Linear Programming}

\end{acronym}

\usepackage{amsmath,amsfonts,bm}

\def\Tableref#1{Table~\ref{#1}}

\def\Figref#1{Figure~\ref{#1}}

\def\Secref#1{Section~\ref{#1}}

\def\eqref#1{equation~\ref{#1}}

\def\1{\bm{1}}

\DeclareMathAlphabet{\mathsfit}{\encodingdefault}{\sfdefault}{m}{sl}
\SetMathAlphabet{\mathsfit}{bold}{\encodingdefault}{\sfdefault}{bx}{n}

\def\gC{{\mathcal{C}}}
\def\gD{{\mathcal{D}}}

\def\gR{{\mathcal{R}}}
\def\gS{{\mathcal{S}}}

\begin{document}
\bstctlcite{BSTcontrol}

\title{Towards Scalable O-RAN Resource Management: Graph-Augmented Proximal Policy Optimization}
\author{
    \IEEEauthorblockN{
        Duc-Thinh Ngo\IEEEauthorrefmark{1}\IEEEauthorrefmark{2}\IEEEauthorrefmark{4}, Kandaraj Piamrat\IEEEauthorrefmark{2}\IEEEauthorrefmark{4}, Ons Aouedi\IEEEauthorrefmark{3},
        Thomas Hassan\IEEEauthorrefmark{1}\IEEEauthorrefmark{4}, Philippe Raipin\IEEEauthorrefmark{1}\IEEEauthorrefmark{4}
    }
    \IEEEauthorblockA{\IEEEauthorrefmark{1} Orange Innovation, Cesson-Sévigné, France}
    \IEEEauthorblockA{\IEEEauthorrefmark{2} Nantes University, École Centrale Nantes, IMT Atlantique, CNRS, INRIA, LS2N, UMR 6004, Nantes, France}
    \IEEEauthorblockA{\IEEEauthorrefmark{3} SnT, SIGCOM, University of Luxembourg, Luxembourg}
}

\maketitle
\begin{abstract}
    Open Radio Access Network (O-RAN) architectures enable flexible, scalable, and cost-efficient mobile networks by disaggregating and virtualizing baseband functions. However, this flexibility introduces significant challenges for resource management, requiring joint optimization of functional split selection and virtualized unit placement under dynamic demands and complex topologies. Existing solutions often address these aspects separately or lack scalability in large and real-world scenarios. In this work, we propose a novel Graph-Augmented Proximal Policy Optimization (GPPO) framework that leverages Graph Neural Networks (GNNs) for topology-aware feature extraction and integrates action masking to efficiently navigate the combinatorial decision space. Our approach jointly optimizes functional split and placement decisions, capturing the full complexity of O-RAN resource allocation. Extensive experiments on both small- and large-scale O-RAN scenarios demonstrate that GPPO consistently outperforms state-of-the-art baselines, achieving up to 18\% lower deployment cost and 25\% higher reward in generalization tests, while maintaining perfect reliability. These results highlight the effectiveness and scalability of GPPO for practical O-RAN deployments.
\end{abstract}
\section{Introduction}

The rapid evolution of mobile networks and the increasing diversity of service requirements have driven the adoption of Open Radio Access Network (O-RAN), which promises greater flexibility, scalability, and cost efficiency through the disaggregation and virtualization of baseband functions. In O-RAN, operators can dynamically select functional splits and flexibly place virtualized baseband units across distributed computing resources, enabling more efficient resource utilization and improved support for heterogeneous \acp{sla}. However, this flexibility introduces significant challenges for resource management, as orchestrators must jointly optimize functional split selection and the placement of virtualized units under dynamic traffic demands, stringent latency constraints, and complex network topologies.

A variety of approaches have been proposed to address these challenges. Early works typically formulate the resource allocation problem as an \acp{ilp} task, focusing on either the optimal placement of baseband units (BU) or the selection of functional splits, and sometimes both \cite{Mharsi2018ScalableCostEfficient,harutyunyanFlexibleFunctionalSplit2017}. More recent studies leverage \acp{drl} to handle the NP-hardness and adapt to dynamic request patterns \cite{murtiConstrainedDeepReinforcement2022, amiriDeepReinforcementLearning2024, murtiDeepReinforcementLearning2024}, while others incorporate \acp{gnn} to better capture the topological relationships in large-scale O-RAN deployments \cite{liNetMindAdaptiveRAN2024}. Despite these advances, most existing solutions either treat functional split and placement as separate problems, assume static or simplified network topologies, or neglect the intricate coupling between routing, resource constraints, and \ac{sla} requirements. As a result, their scalability and effectiveness in real-world, large-scale, and dynamic O-RAN environments remain limited.

In this work, we propose a novel, scalable framework for O-RAN resource management that jointly optimizes functional split selection and virtualized baseband unit placement. Our approach leverages a Graph-Augmented Proximal Policy Optimization (GPPO) agent, which integrates \ac{gnn}-based feature extraction with advanced DRL techniques and action masking to efficiently navigate the combinatorial decision space. By modeling the O-RAN substrate as a graph and encoding both node and edge attributes, our method captures the full complexity of network topology, resource availability, and service demands. Extensive experiments on both small- and large-scale O-RAN scenarios demonstrate that our GPPO approach consistently outperforms state-of-the-art baselines in terms of deployment cost, reward, and reliability, achieving superior scalability and solution quality even in highly dynamic and complex environments. Comparing to the closest works \cite{murtiDeepReinforcementLearning2024,liNetMindAdaptiveRAN2024}, our method highlights the joint optimization of functional split and placement, while incorporating \ac{gnn} to model the complex relationships in the O-RAN substrate, leading to more scalable and efficient resource allocation decisions.

The rest of the paper is organized as follows: Section \ref{sec:related} reviews related works on O-RAN resource allocation, highlighting the limitations of existing approaches. Section \ref{sec:sm} defines the system model, and Section \ref{sec:pf} formulates the O-RAN resource allocation problem and optimization objectives. Section \ref{sec:method} presents our GPPO framework, detailing the GNN-based feature extraction, PPO training process, and action masking techniques. Section \ref{sec:exp} describes the experimental setup and results, comparing our method against state-of-the-art baselines. Finally, Section \ref{sec:conclusion} concludes the paper and discusses future work.
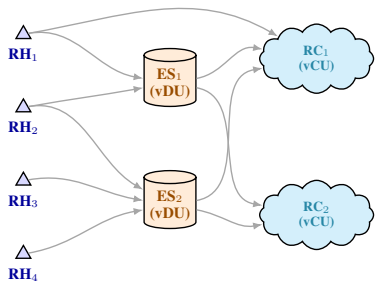
\begin{figure}
    \centering
    \resizebox{0.6\linewidth}{!}{
        \resizebox{\linewidth}{!}{
\begin{tikzpicture}[
    rh/.style={draw, regular polygon, regular polygon sides=3, fill=blue!15, minimum size=0.35cm, inner sep=0pt, line width=0.8pt, shape border rotate=120},
    es/.style={draw, cylinder, shape border rotate=90, aspect=0.25, minimum height=1cm, minimum width=0.7cm, fill=orange!15, font=\bfseries\small, text=orange!60!black, line width=0.8pt},
    rc/.style={draw, cloud, cloud puffs=12, cloud puff arc=120, aspect=2, fill=cyan!15, minimum width=1.2cm, minimum height=1.1cm, font=\bfseries\footnotesize, text=cyan!60!black, line width=0.8pt},
    link/.style={thick, -{Latex[length=2mm]}, color=gray!70},
    every node/.style={align=center}
]
    \node[rh] (rh1) at (0,2.5) {};
    \node[below=2pt of rh1, text=blue!60!black, font=\bfseries\small, align=center] {RH$_1$};
    \node[rh] (rh2) at (0,1) {};
    \node[below=2pt of rh2, text=blue!60!black, font=\bfseries\small, align=center] {RH$_2$};
    \node[rh] (rh3) at (0,-0.5){};
    \node[below=2pt of rh3, text=blue!60!black, font=\bfseries\small, align=center] {RH$_3$};
    \node[rh] (rh4) at (0,-2)  {};
    \node[below=2pt of rh4, text=blue!60!black, font=\bfseries\small, align=center] {RH$_4$};

    \node[es] (es1) at (3,1.5) {ES$_1$\\(vDU)};
    \node[es] (es2) at (3,-1)  {ES$_2$\\(vDU)};

    \node[rc] (rc1) at (6,2)   {RC$_1$\\(vCU)};
    \node[rc] (rc2) at (6,-1.2){RC$_2$\\(vCU)};

    \draw[link] (rh1) to[out=10,in=170] (es1);
    \draw[link] (rh2) to[out=10,in=190] (es1);
    \draw[link] (rh2) to[out=10,in=150] (es2);
    \draw[link] (rh3) to[out=10,in=170] (es2);
    \draw[link] (rh4) to[out=10,in=190] (es2);

    \draw[link] (es1) to[out=10,in=170] (rc1);
    \draw[link] (es1) to[out=350,in=170] (rc2);
    \draw[link] (es2) to[out=10,in=190] (rc1);
    \draw[link] (es2) to[out=350,in=190] (rc2);

    \draw[link] (rh1) to[out=10,in=150] (rc1);

\end{tikzpicture}
}
    }
    \caption{O-RAN flexible functional split and placement architecture.}
    \label{fig:oran_flex_split}
    \vspace{-0.4cm}

\end{figure}

\section{Related works}\label{sec:related}
\vspace{-0.2cm}

We investigate the related works on O-RAN resource allocation, focusing on the following aspects: functional split selection, baseband unit placement, dynamic requests, routing, and deep learning approaches. A summary of the related works is provided in \Tableref{tab:compare_related_work}.

\subsection{Joint optimization of functional split and BU placement}
We can decompose the resource allocation problem in \ac{ran} into two lines of subproblems: baseband unit placement and optimization of flexible functional split. The former addresses the efficient deployment of baseband processing units across computing nodes in centralized or virtualized RAN architectures. For instance, \cite{Mharsi2018ScalableCostEfficient,aliProactiveVNFScaling2024,liNetMindAdaptiveRAN2024} focus uniquely on optimizing the dynamic placement of virtual baseband units on physical DUs and CUs in order to minimize the energy and consumption rate. In contrast, the latter leverages the flexibility of O-RAN base stations to dynamically select the most suitable functional split for each station, based on their specific service demands, often driven by the requirements of different network slices. However, these works often accompany the optimization of virtualized baseband unit placement, as the two problems are closely related. For example, \cite{moraisPlaceRANOptimalPlacement2022,murtiConstrainedDeepReinforcement2022} focus on optimizing the functional split selection and vDU placement, while more recent works \cite{lopesORANOrientedApproachDynamic2024,amiriDeepReinforcementLearning2024,murtiDeepReinforcementLearning2024} consider vCU placement additionally. Overall, these problems can always be formulated as an \ac{ilp} problem, which is NP-hard.

\subsection{Dynamic requests}
The dynamic nature of requests in O-RAN systems introduces additional complexity to the resource allocation problem. It can incur significant overhead in terms of reconfiguration and resource management, as the system must adapt to changing traffic patterns and service requirements. Most of the existing works focus on proposing a single configuration given static requests. For instance, \cite{moraisPlaceRANOptimalPlacement2022,murtiConstrainedDeepReinforcement2022} consider solving an optimization problem for each timeslot independently, which does not account for the dynamic nature of requests. In contrast, \cite{lopesORANOrientedApproachDynamic2024,murtiDeepReinforcementLearning2024} propose solutions to the problem of reconfiguration that can handle dynamic requests, allowing for more efficient resource allocation and management.

\subsection{Routing}
In a large-scale O-RAN architecture with complex topology, routing, which is a coupling problem with the resource placement problem, is crucial and challenging to address. Moreover, when routing is considered in a large-scale substrate network, it can increase the complexity of the problem combinatorially. Most of the existing works focus on the placement of virtualized baseband units and functional split selection, while routing is often assumed to be static or predefined. For example, \cite{moraisPlaceRANOptimalPlacement2022} considers a single-CU scenario, where the routing is straightforward as all \acp{rh} are connected to a single CU. \cite{murtiConstrainedDeepReinforcement2022} considered that each \ac{rh} is physically associated to a distinct DU, neglecting the need to optimize the routing policy. In \cite{lopesORANOrientedApproachDynamic2024}, CU placement is decided by the optimizer but the routing (choosing DU placement in this case) is decided by the shortest path algorithm, which only partially aligns with the cost minimization objective. Recent works accounting for routing, e.g., \cite{liNetMindAdaptiveRAN2024,murtiDeepReinforcementLearning2024} suffer from seriously limited scalability, leading them to only consider small-scale networks with a limited number of nodes. In our approach, we propose a novel method that integrates routing optimization into the resource allocation process but still maintains scalability, allowing for efficient routing decisions even in large-scale O-RAN networks.

\subsection{Deep learning approaches}
To address the NP-hard and dynamic nature of the O-RAN resource allocation problem, many existing works leverage deep learning techniques, particularly \ac{drl} \cite{murtiConstrainedDeepReinforcement2022, amiriDeepReinforcementLearning2024,murtiDeepReinforcementLearning2024,liNetMindAdaptiveRAN2024}. These approaches can learn by automatically exploring and exploiting the simulated environment to make optimal decisions based on the current state of the system. Moreover, to resolve the challenging routing problem, especially when the topology is complex, \acp{gnn} can be used to model the relationships between nodes and edges in the network, allowing for more efficient allocation decisions \cite{liNetMindAdaptiveRAN2024}. In this work, we propose a novel approach that combines \ac{drl} with \acp{gnn} to address the O-RAN resource allocation problem, leveraging the strengths of both techniques to achieve both scalability and efficiency.

\begin{table}
    \centering
    \begin{threeparttable}
        \setlength{\tabcolsep}{4pt}
        \caption{Summary of related works on RAN resource allocation.}
        \label{tab:compare_related_work}
        \begin{tabular}{@{}cccccccc@{}}
            Ref. & \makecell{Flexible\\split} & \makecell{VNF\\placement} & Routing & \makecell{Scale*} & \makecell{Dynamic\\requests} & \makecell{DRL} & \makecell{GCN}\\ \midrule
            \cite{moraisPlaceRANOptimalPlacement2022} & \checkmark &  &  & 128 &  &  & \\
            \cite{murtiConstrainedDeepReinforcement2022} & \checkmark &  &  & 201 &  & \checkmark &  \\ 
            \cite{lopesORANOrientedApproachDynamic2024} & \checkmark & \checkmark &  & 21 & \checkmark & \checkmark &  \\
            \cite{liNetMindAdaptiveRAN2024} &  & \checkmark & \checkmark & 6 & \checkmark & \checkmark & \checkmark \\
            \cite{murtiDeepReinforcementLearning2024} & \checkmark & \checkmark & \checkmark & 20 & \checkmark & \checkmark &  \\ \midrule
            Ours & \checkmark & \checkmark & \checkmark & 70 & \checkmark & \checkmark & \checkmark
        \end{tabular}
        \begin{tablenotes}
            \item *: Scale of the problem, defined as the number of nodes in the network.
        \end{tablenotes}
        \vspace{-0.4cm}

    \end{threeparttable}
\end{table}

\section{System model}\label{sec:sm}

We consider an O-RAN architecture comprising three types of nodes: \textit{\acp{rh}}, \textit{\acp{es}}, and \textit{\acp{rc}}. The baseband processing functions are disaggregated and distributed across vDUs hosted on \acp{es} and vCUs hosted on \acp{rc}, connected through the cross-haul network. \Figref{fig:oran_flex_split} illustrates an example of a small hierarchical O-RAN. At each time slot, under a uniform temporal granularity, a request arrives consisting of a traffic load and an associated end-to-end latency requirement. These requests are categorized into three primary slice types, each associated with distinct \acp{sla}:\textit{ \ac{embb}}, \textit{\ac{urllc}}, and \textit{\ac{mmtc}}. For every \ac{rh}, the O-RAN orchestrator must simultaneously make two key decisions: (i) select an appropriate functional split and (ii) allocate virtualized baseband processing units to computing nodes.

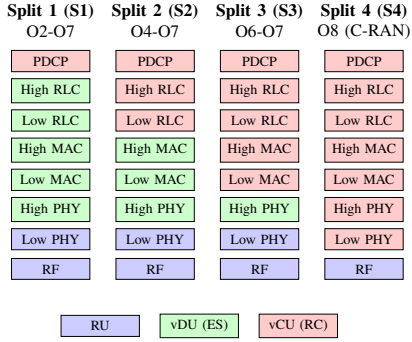
\begin{figure}
    \centering
    \resizebox{0.65\linewidth}{!}{

\begin{tikzpicture}[
    layer/.style={draw, rectangle, minimum width=45pt, minimum height=10pt, align=center, font=\footnotesize},
    ru/.style={layer, fill=blue!20},
    du/.style={layer, fill=green!20},
    cu/.style={layer, fill=red!20},
    split/.style={draw, thick, red, dashed},
    title/.style={font=\bfseries},
    subtitle/.style={font=}
]

\begin{scope}[xshift=0pt]
    \node[title] at (0, 5.5) {Split 1 (S1)};
    \node[subtitle] at (0, 5.1) {O2-O7};
    
    \node[ru] at (0, 0.3) {RF};
    \node[ru] at (0, 0.9) {Low PHY};
    
    \node[du] at (0, 1.5) {High PHY};
    \node[du] at (0, 2.1) {Low MAC};
    \node[du] at (0, 2.7) {High MAC};
    \node[du] at (0, 3.3) {Low RLC};
    \node[du] at (0, 3.9) {High RLC};
    
    \node[cu] at (0, 4.5) {PDCP};
    
    
\end{scope}

\begin{scope}[xshift=60pt]
    \node[title] at (0, 5.5) {Split 2 (S2)};
    \node[subtitle] at (0, 5.1) {O4-O7};
    
    \node[ru] at (0, 0.3) {RF};
    \node[ru] at (0, 0.9) {Low PHY};
    
    \node[du] at (0, 1.5) {High PHY};
    \node[du] at (0, 2.1) {Low MAC};
    \node[du] at (0, 2.7) {High MAC};
    
    \node[cu] at (0, 3.3) {Low RLC};
    \node[cu] at (0, 3.9) {High RLC};
    \node[cu] at (0, 4.5) {PDCP};
    
    
\end{scope}

\begin{scope}[xshift=120pt]
    \node[title] at (0, 5.5) {Split 3 (S3)};
    \node[subtitle] at (0, 5.1) {O6-O7};
    
    \node[ru] at (0, 0.3) {RF};
    \node[ru] at (0, 0.9) {Low PHY};
    
    \node[du] at (0, 1.5) {High PHY};
    
    \node[cu] at (0, 2.1) {Low MAC};
    \node[cu] at (0, 2.7) {High MAC};
    \node[cu] at (0, 3.3) {Low RLC};
    \node[cu] at (0, 3.9) {High RLC};
    \node[cu] at (0, 4.5) {PDCP};
    
    
\end{scope}

\begin{scope}[xshift=180pt]
    \node[title] at (0, 5.5) {Split 4 (S4)};
    \node[subtitle] at (0, 5.1) {O8 (C-RAN)};
    
    \node[ru] at (0, 0.3) {RF};
    
    \node[cu] at (0, 0.9) {Low PHY};
    \node[cu] at (0, 1.5) {High PHY};
    \node[cu] at (0, 2.1) {Low MAC};
    \node[cu] at (0, 2.7) {High MAC};
    \node[cu] at (0, 3.3) {Low RLC};
    \node[cu] at (0, 3.9) {High RLC};
    \node[cu] at (0, 4.5) {PDCP};
    
    
\end{scope}

\begin{scope}[yshift=-10pt]
    \node[ru] at (1, -0.5) {RU};
    \node[du] at (3, -0.5) {vDU (ES)};
    \node[cu] at (5, -0.5) {vCU (RC)};
\end{scope}

\end{tikzpicture}
    }\vspace{-0.2cm}

    \caption{Functional split configurations showing protocol layer distribution across RU, vDU (hosted on ES), and vCU (hosted on RC). 
    }
    \label{fig:functional_splits}
    \vspace{-0.4cm}
\end{figure}

Following standardized O-RAN recommendations~\cite{murtiDeepReinforcementLearning2024}, we consider four primary functional split configurations that combine high-layer splits (HLS) between vCU and vDU with low-layer splits (LLS) between vDU and radio units as shown in Figure \ref{fig:functional_splits}. 
In practical deployments, these constraints are multifaceted and include both feasibility conditions and \ac{sla}-driven requirements. From a feasibility standpoint, the placement of vDUs and vCUs must ensure connectivity between the selected \ac{es}, \ac{rc}, and their associated \ac{rh}, while also satisfying the technical requirements imposed by the selected functional split —particularly regarding cross-haul latency and bandwidth. Each split imposes distinct cross-haul bandwidth and latency requirements, with more centralized splits demanding higher bandwidth capacity and stricter latency bounds. To avoid overloading compute resources and to meet \ac{sla} targets, the orchestrator must also ensure that server capacity constraints are not violated and that the end-to-end latency remains within the required bounds.

Beyond maximizing the success ratio, deployment cost is a critical optimization objective. These costs encompass both processing and routing fees. In O-RAN systems, processing on \acp{es} typically incurs higher costs than on \acp{rc}, which incentivizes centralizing baseband functions at regional clouds. However, such centralization may increase routing costs or, in some cases, prevent \ac{sla} compliance due to tighter constraints on bandwidth and latency.

\begin{table}
    \centering
    \begin{threeparttable}
        \centering
        \caption{Functional split options and their associated traffic load and delay requirements.}
        \label{tab:placeholder_label}
        \begin{tabular}{@{}lcc@{}}
            \toprule
            \textbf{Split option} & \textbf{Traffic load (Gbps)} & \textbf{Delay (ms)} \\ \midrule
            O2 & $\lambda$ & 10 \\ 
            O4 & $\lambda$ & 1 \\
            O6 & $1.02 \lambda + 0.5$ & 0.25 \\
            O7 & 10.01 & 0.25 \\
            O8 & 157.3 & 0.25 \\ \bottomrule
        \end{tabular}
        \begin{tablenotes}
            \item Note: $\lambda$ is the traffic load of the \ac{rh} in Gbps, which is determined by the service slice type.
        \end{tablenotes}
        \vspace{-0.6cm}
    \end{threeparttable}
\end{table}

\section{Problem formulation} \label{sec:pf}

In this section, we formulate the dynamic O-RAN resource allocation problem as a multi-objective optimization problem that jointly addresses functional split selection and virtualized baseband unit placement. The optimization decisions must be made at each time slot $t$ while satisfying multiple constraints related to resource capacity, connectivity, and \ac{sla} requirements. \vspace{-0.7cm}

\subsection{Problem Scope and Formulation}


\begin{table}[!t]
\centering
\footnotesize
\caption{Summary of Notation}
\vspace{-0.2cm}
\label{tab:notation}
\setlength{\tabcolsep}{3pt}
\begin{tabular}{@{}cl@{}}
\toprule
\textbf{Symbol} & \textbf{Description} \\
\midrule
\multicolumn{2}{l}{\textit{Sets and Indices}} \\
$\gR$ & Set of \acp{rh}, indexed by $r$ \\
$\gD$ & Set of \acp{es}, indexed by $d$ \\
$\gC$ & Set of \acp{rc}, indexed by $c$ \\
$\gS$ & Set of functional split options, indexed by $s$ \\
$t$ & Time slot index \\
\midrule
\multicolumn{2}{l}{\textit{Decision Variables}} \\
$f^t_{rs}$ & Binary: \ac{rh} $r$ uses split $s$ at time $t$ \\
$x^t_{rd}$ & Binary: vDU of \ac{rh} $r$ placed on \ac{es} $d$ at time $t$ \\
$y^t_{rc}$ & Binary: vCU of \ac{rh} $r$ placed on \ac{rc} $c$ at time $t$ \\
\midrule
\multicolumn{2}{l}{\textit{Network Parameters}} \\
$e_{dc}$ & Binary: connectivity between \ac{es} $d$ and \ac{rc} $c$ \\
$\delta_{rd}$ & Link delay between \ac{rh} $r$ and \ac{es} $d$ \\
$\delta_{dc}$ & Link delay between \ac{es} $d$ and \ac{rc} $c$ \\
$B_{dc}$ & Bandwidth capacity of link between \ac{es} $d$ and \ac{rc} $c$ \\
\midrule
\multicolumn{2}{l}{\textit{Resource Parameters}} \\
$c^{\rm DU}_s$ & Computing resource requirement for vDU under split $s$ \\
$c^{\rm CU}_s$ & Computing resource requirement for vCU under split $s$ \\
$R_d$ & Computing capacity of \ac{es} $d$ \\
$R_c$ & Computing capacity of \ac{rc} $c$ \\
$\beta^s$ & Traffic load generated by split $s$ \\
\midrule
\multicolumn{2}{l}{\textit{QoS Parameters}} \\
$\Delta_r$ & End-to-end latency requirement for \ac{rh} $r$ \\
$\Delta^{\rm DU-CU}_s$ & Cross-haul latency requirement for split $s$ \\
\midrule
\multicolumn{2}{l}{\textit{Cost Parameters}} \\
$\psi^{\rm DU}_s$ & Unit cost for vDU deployment of split $s$ \\
$\psi^{\rm CU}_s$ & Unit cost for vCU deployment of split $s$ \\
$\phi_{\rm R}$ & Reconfiguration penalty factor \\
$\phi_{\rm L}$ & Routing cost factor \\
\midrule
\multicolumn{2}{l}{\textit{Objective Functions}} \\
$J_{\rm C}$ & Computing resource cost \\
$J_{\rm R}$ & Reconfiguration cost \\
$J_{\rm L}$ & Routing cost \\
\bottomrule
\end{tabular}\vspace{-0.6cm}

\end{table}

We consider a discrete-time system where decisions are made at each time slot $t \in \{1, 2, \ldots, T\}$. At each time slot, the O-RAN orchestrator must jointly determine: (i) the functional split assignment for each \ac{rh}, (ii) the placement of vDUs on \acp{es}, and (iii) the placement of vCUs on \acp{rc}. These decisions are represented by the binary decision variables $f^t_{rs}$, $x^t_{rd}$, and $y^t_{rc}$ as defined in \Tableref{tab:notation}. 
The optimization problem aims to minimize the total cost while satisfying resource capacity, connectivity, and \ac{sla} constraints. 
The optimization problem is subject to the following constraints:

\subsubsection{Assignment Constraints}
Each \ac{rh} must be assigned exactly one functional split, one \ac{es} for vDU placement, and one \ac{rc} for vCU placement:\vspace{-0.2cm}
\begin{align}
    \sum_{s \in \gS} f^t_{rs} &= 1, \quad \forall r \in \gR, t \tag{C1}\label{eq:split_assign}\\
    \sum_{d \in \gD} x^t_{rd} &= 1, \quad \forall r \in \gR, t \tag{C2}\label{eq:du_assign}\\
    \sum_{c \in \gC} y^t_{rc} &= 1, \quad \forall r \in \gR, t \tag{C3}\label{eq:cu_assign}
\end{align}

\subsubsection{Connectivity Constraint}
The selected \ac{es} and \ac{rc} for each \ac{rh} must be connected by a physical link:\vspace{-0.2cm}
\begin{align}
    x^t_{rd}y^t_{rc} \leq e_{dc}, \quad \forall r \in \gR, d \in \gD, c \in \gC, t \tag{C4}\label{eq:conn}
\end{align}

\subsubsection{Resource Capacity Constraints}
The total computing resource consumption on each server cannot exceed its capacity:\vspace{-0.2cm}
\begin{align}
    p_d = \sum_{r\in \gR} x^t_{rd} \sum_{s \in\gS}c^{\rm DU}_s f^t_{rs} &\leq R_d, \quad \forall d \in \gD, t \tag{C5}\label{eq:es_capacity}\\
    p_c = \sum_{r\in \gR} y^t_{rc} \sum_{s \in\gS}c^{\rm CU}_s f^t_{rs} &\leq R_c, \quad \forall c \in \gC, t \tag{C6}\label{eq:rc_capacity}
\end{align}

\subsubsection{End-to-End Latency Constraint}
The total end-to-end delay from each \ac{rh} to its assigned \ac{rc} must satisfy the \ac{sla} requirement:\vspace{-0.2cm}
\begin{align}
    l_{\rm e2e}^r = \sum_{d \in \gD, c \in \gC} (\delta_{rd} + \delta_{dc}) x^t_{rd} y^t_{rc} \leq \Delta_r, \quad \forall r \in \gR, t \tag{C7}\label{eq:e2e_delay}
\end{align}

\subsubsection{Cross-Haul Latency Constraint}
The cross-haul delay between vDU and vCU must not exceed the requirement imposed by the selected functional split:\vspace{-0.2cm}
\begin{align}
    l^{\rm cross}_r = \sum_{d \in \gD, c \in \gC} \delta_{dc} x^t_{rd} y^t_{rc} \leq \sum_{s\in \gS}\Delta^{\rm DU-CU}_s f^t_{rs}, \quad \forall r \in \gR, t \tag{C8}\label{eq:cross_delay}
\end{align}

\subsubsection{Link Bandwidth Constraint}
The total traffic on each cross-haul link must not exceed its capacity:\vspace{-0.2cm}
\begin{align}
    b_{dc} = \sum_{r \in \gR} x^t_{rd} y^t_{rc} \sum_{s \in \gS}\beta^s f^t_{rs} \leq B_{dc}, \quad \forall d \in \gD, c \in \gC, t \tag{C9}\label{eq:bandwidth}
\end{align}

\subsection{Objective Function}

The optimization problem aims to minimize the total deployment cost, which comprises three components:

\subsubsection{Computing Resource Cost}
This component captures the cost of deploying vDUs and vCUs, where processing on \acp{es} is more expensive than on \acp{rc}:
\begin{equation}
\begin{split}
    J_{\rm C} &= \sum_{r\in \gR, d\in \gD} x^t_{rd} \sum_{s \in\gS}c^{\rm DU}_s f^t_{rs} \psi^{\rm DU}_s\\
    &\quad + \sum_{r\in \gR, c\in \gC} y^t_{rc} \sum_{s \in\gS}c^{\rm CU}_s f^t_{rs}\psi^{\rm CU}_s
\end{split}
\label{eq:compute_cost}
\end{equation}
where $\psi^{\rm DU}$ and $\psi^{\rm CU}$ are the unit costs for vDU and vCU deployment, respectively.

\subsubsection{Reconfiguration Cost}
To account for the operational overhead of changing network configurations, we include a penalty for placement modifications between consecutive time slots:\vspace{-0.4cm}
\begin{align}
    J_{\rm R} = \phi_{\rm R} \left(\sum_{r,d}\left\lvert x^{t+1}_{rd} - x^t_{rd} \right\rvert + \sum_{r,c}\left\lvert y^{t+1}_{rc} - y^t_{rc} \right\rvert\right)
    \label{eq:reconfig_cost}
\end{align}
where $\phi_{\rm R}$ is the reconfiguration penalty factor.

\subsubsection{Routing Cost}
This component represents the cost of utilizing network links for cross-haul traffic:
\begin{align}
    J_{\rm L} = \phi_{\rm L} \sum_{d \in \gD,c \in \gC } \delta_{dc} \sum_{r \in \gR} x^t_{rd} y^t_{rc} \sum_{s \in \gS}\beta^s f^t_{rs}
    \label{eq:routing_cost}
\end{align}
where $\phi_{\rm L}$ is the routing cost factor, and the term $\sum_{r \in \gR} x^t_{rd} y^t_{rc} \sum_{s \in \gS}\beta^s f^t_{rs}$ represents the total traffic on link $(d,c)$.

\subsubsection{Total Objective}
The complete optimization problem is formulated as: \vspace{-0.4cm}
\begin{align}
    \min \quad & J_{\rm C} + J_{\rm R} + J_{\rm L} \label{eq:total_objective}\\
    \text{s.t.} \quad & \text{Constraints (\ref{eq:split_assign}--\ref{eq:bandwidth})} \nonumber
\end{align}
This formulation balances efficient resource utilization, system stability, and overall network performance while enforcing all SLA and capacity constraints. 
Because all decision variables are binary and constraints are linear, the problem is \ac{ilp}-compliant. 
However, it remains NP-hard \cite{murtiConstrainedDeepReinforcement2022, amiriDeepReinforcementLearning2024}, making exact optimization in large networks impractical. 
Therefore, we employ a reinforcement learning approach to efficiently explore the combinatorial decision space and learn effective placement and split-selection policies.

\section{Methodology}\label{sec:method}
\begin{figure}
    \centering
    \resizebox{0.7\linewidth}{!}{
        \resizebox{\linewidth}{!}{
\begin{tikzpicture}[
    node distance=1.5cm and 2.5cm,
    every node/.style={font=},
    env/.style={draw, rectangle, rounded corners, fill=blue!15, minimum width=2.8cm, minimum height=1.1cm},
    process/.style={draw, rectangle, rounded corners, fill=orange!15, minimum width=2.5cm, minimum height=1cm},
    trainable/.style={draw, rectangle, rounded corners, fill=green!15, minimum width=2.5cm, minimum height=1cm, align=center},
    action/.style={draw, rectangle, rounded corners, fill=red!15, minimum width=2.5cm, minimum height=1cm},
    thickarrow/.style={thick, -{Latex[length=2mm]}},
    ppo/.style={draw, rectangle, rounded corners, dashed, thick, inner sep=0.3cm, minimum width=3.5cm, minimum height=3.2cm, label={[font=\bfseries\footnotesize, fill=white]0:PPO}}
  ]
  \node[env] (env) at (0,2) {O-RAN Environment};
  \node[trainable] (extractor) at (0,-0.5) {GNN-based\\Feature Extractor};
  \node[trainable, below=of extractor] (policy) {Policy \& Value\\Networks};
  \node[action, right=1.2cm of policy] (actionmask) {Action Mask};
  \node[action, above=of actionmask] (action) {Action};
  \node[process, left=1.8cm of policy, yshift=1.25cm] (reward) {Reward};

  \begin{pgfonlayer}{background}
    \node[ppo, fit=(extractor) (policy)] (ppobox) {};
  \end{pgfonlayer}

  \draw[thickarrow] (env) -- node[right, align=left] {Generate\\next state} (extractor);
  \draw[thickarrow] (env) -| node[pos=0.05,above left] {Compute Reward} (reward);
  \draw[thickarrow] (extractor) -- (policy);
  \draw[thickarrow] ([yshift=0cm]reward.east) --node[above] {Update} (ppobox.west);
  \draw[thickarrow] (policy) -- (actionmask);
  \draw[thickarrow] (actionmask) -- (action);
  \draw[thickarrow] (action) |- node[pos=0.85,above right] {Apply Action} (env);

\end{tikzpicture}
}
    }
    \caption{Workflow of the GNN-enhanced PPO agent for O-RAN placement.}\vspace{-0.4cm}
    \label{fig:method_overview}
\end{figure}
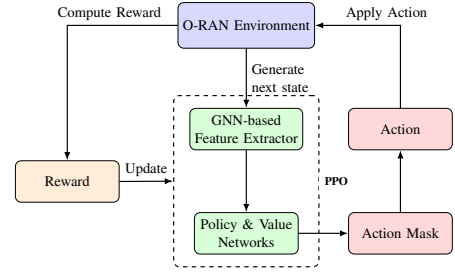

\subsection{O-RAN placement environment}

To be able to solve this dynamic placement problem, we first reformulate it into an \ac{mdp}, represented by a 4-tuple $(S, A, P_a, R_a)$. $S$ and $A$ are the space of states and actions. $P_a(s^\prime, s)$ is the probability of the transition from state $s^\prime$ to the state $s$, conditional on the taken action $a$. Similarly, $R_a(s^\prime, s)$ is the immediate reward when the environment state transists from $s^\prime$ to $s$ in the effect of the taken action $a$. While the transition probability is usually unknown and has to be implicitly learned from the exploration of the RL agent, the state-action space and reward function have to be explicitly defined when constructing an RL environment. To better reflect the optimization problem, we construct an \ac{mdp} as follows:

\subsubsection{State}
The state variables will contain information about the current status of the environment, allowing the RL agent to reason and take the proper actions. Therefore, the state space should provide information about the demands of each \ac{rh} and remaining computing resources of servers, remaining bandwidth of all links, and other static information, i.e., substrate network topology and link delays.
\subsubsection{Action}
The action space is the sample space of all actions that an RL agent can take. In this joint optimization problem, the RL agent has to make decisions on functional split assignment and virtual units placement simultaneously. Therefore, the action vector must be a vector of dimension $3N$ where the first $N$ discrete variables are to indicate which functional split should be assigned to the corresponding \ac{rh}. Besides, the last $2N$ are discrete variables to indicate which \ac{es} and \ac{rc} should be the hosts of the corresponding \ac{rh}'s vDU and vCU. 

Formally, we denote the action vector at time $t$ as:
\begin{align}
    \mathbf{a}_t = [a^{\text{split}}_{1,t}, \ldots, a^{\text{split}}_{N,t},\ a^{\text{DU}}_{1,t}, \ldots, a^{\text{DU}}_{N,t},\ a^{\text{CU}}_{1,t}, \ldots, a^{\text{CU}}_{N,t}]
\end{align}
where $a^{\text{split}}_{i,t}$ indicates the functional split selected for \ac{rh} $i$, $a^{\text{DU}}_{i,t}$ indicates the selected \ac{es} (vDU placement) for \ac{rh} $i$, and $a^{\text{CU}}_{i,t}$ indicates the selected \ac{rc} (vCU placement) for \ac{rh} $i$.
\subsubsection{Reward}
While the original problem is to minimize the cost, the MDP's natural objective is to maximize the rewards. Moreover, due to the large action space, which grows exponentially with the number of \ac{rh}, the exploration process of the RL agent becomes much more difficult since the number of invalid actions also increases exponentially.

One solution to resolve this sparse reward problem is to design the reward system with care, so that even when invalid actions are taken, the returned reward can give the agent hints about the direction of the exploration, thus guiding the exploration more efficiently.

Another solution is to relax constraints. We notice that among the constraints raised in \Secref{sec:pf}, only the constraint \ref{eq:conn} concerns the feasibility of deployment. Other constraints are mostly \ac{sla}-driven. Therefore, even if these constraints are violated, the deployment is still feasible. Thus, we can relax these constraints and impose a cost of \ac{sla}-violation instead. This cost will be added to the reward function we are constructing. \vspace{-0.4cm}

\begin{align}
    J_{\rm SLA} = \sum_{i \in \gD \cup \gC}\max(R_i - p_i, 0) + \sum_{r \in \gR} \max(\Delta_r - l^{\rm e2e}, 0) \notag \\
    + \sum_{r \in \gR} \max(\Delta_r - l^{\rm cross}, 0) + \sum_{i \in \gD \cup \gC}\max(R_i - p_i, 0)
\end{align}
Thus, the total cost in a relaxed optimization problem is:
\begin{align}
    J = J_{\rm L} + J_{\rm R} + J_{\rm C} + J_{\rm SLA}
\end{align}
This allows us to construct our reward function as:
\begin{align}
    R &= \begin{cases}
    -\frac{n_{\rm fail}}{2N}, &\text{if \ref{eq:conn} not satisfied $\forall d, c$}.\\
    -1, &\text{if early terminated.} \\
    (1 + \log{(1 + J)})^{-1}, &\text{otherwise}.
  \end{cases}
\end{align}
where $n_{\rm fail}$ is the number of failed links, which is always smaller than the total of established links $2N$, since each \ac{rh} requires at most 2 links RU-DU and DU-CU. This reward function is ensured to be bounded in $[-1, 1]$. When the action is infeasible, which does not meet the constraint \ref{eq:conn}, the reward is negative, indicating a penalty. When the action is feasible, the reward is positive and is a decreasing function of the deployment cost, meaning that maximizing the reward leads to minimizing the cost.

In our environment, an episode is early terminated if the agent takes 5 consecutive invalid actions (i.e., actions that violate the hard connectivity constraint~(C4)). When this occurs, the environment ends the episode immediately and assigns a reward of $-1$ for that step. This mechanism discourages the agent from repeatedly exploring infeasible regions of the action space and accelerates learning by providing immediate negative feedback for persistent constraint violations.

\subsection{\acf{ppo} with Invalid Action Masking}
To solve the formulated \ac{mdp}, we adopt the \ac{ppo} \cite{schulman2017ppo} algorithm, a widely used on-policy reinforcement learning method that offers a good trade-off between stability and performance. \Ac{ppo} is designed to improve learning stability by preventing the policy from changing too drastically during each update. Given a policy parameterized by $\theta$, the \ac{ppo} objective is defined as:\vspace{-0.4cm}

\begin{equation}
L^{\text{CLIP}}(\theta) = \mathbb{E}_t \left[ \min\left( r_t(\theta) \hat{A}_t, \ \text{clip}(r_t(\theta), 1 - \epsilon, 1 + \epsilon) \hat{A}_t \right) \right]
\end{equation}
where $r_t(\theta) = \frac{\pi_\theta(a_t \mid s_t)}{\pi_{\theta_{\text{old}}}(a_t \mid s_t)}$ is the probability ratio between the new and old policies, and $\hat{A}_t$ is the estimated advantage at time $t$.

The key idea behind \ac{ppo} is to optimize a surrogate objective function that includes a clipping mechanism. This mechanism limits how much the new policy can differ from the previous one during training. Instead of allowing the policy to fully follow the estimated advantage of each action (which can be noisy), \ac{ppo} clips updates that go too far, ensuring more gradual learning and avoiding instability. Compared to the earlier method TRPO \cite{schulman2015trust}, \ac{ppo} achieves similar stability without requiring complex second-order optimization techniques, making it faster and easier to implement.

In our problem, the joint decision space includes functional split selection and vDU/vCU placement for every \ac{rh}. As the number of \acp{rh} increases, the action space grows combinatorially, resulting in a large number of invalid or suboptimal actions, particularly those that violate hard feasibility constraints such as broken connectivity between \ac{es} and \ac{rc} nodes.

To improve learning efficiency and avoid wasting exploration steps on infeasible actions, we integrate action masking into \ac{ppo}~\cite{huang2020closer}. The agent is then restricted to sampling actions only from the valid subset defined by the current action mask. This will prevent the agent from exploring unnecessary invalid actions and thus accelerate the training progress. In this solution, we apply a mask to the action logits before applying the SoftMax function so that the probability of the masked action is reduced to 0. It has been proved that this masking still produces a valid policy gradient~\cite{huang2020closer}.

In our problem, we mask out all $a^{\text{DU}}_{i,t}$ and $a^{\text{CU}}_{i,t}$ actions that are not connected to the \ac{rh} $i$ in the substrate network, i.e., all $d \in \gD$ and $c \in \gC$ such that $e_{rd} = 0$ and $\sum_{d \in \gD} e_{rd} e_{dc} = 0$. This ensures that the agent only considers feasible placements that maintain connectivity between the vDU and vCU of each \ac{rh}. However, this does not guarantee that the proposed placement is feasible, as the selected \ac{rc} are not guaranteed to be connected to the selected \ac{es}. Similarly, we mask out all $a^{\text{split}}_{i,t}=4$ (Split 4) actions for \acp{rh} that do not have a direct link to any \ac{rc} in the substrate network, i.e., all $c \in \gC$ such that $e_{rc} = 0$, as Split 4 requires a direct connection between the \ac{rh} and \ac{rc}.

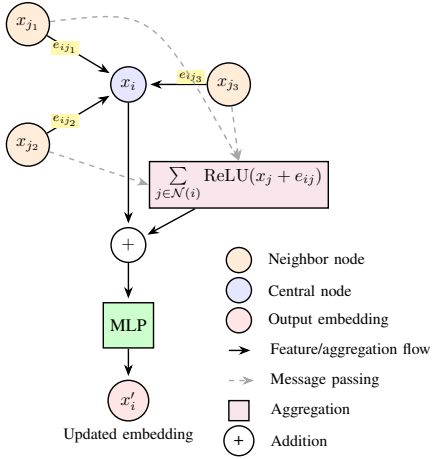
\begin{figure}
    \centering
    \resizebox{0.7\linewidth}{!}{
        \resizebox{0.8\linewidth}{!}{
\begin{tikzpicture}[
    node/.style={circle, draw, fill=blue!10, minimum size=0.5cm, font=, thick},
    neighbor/.style={circle, draw, fill=orange!15, minimum size=0.5cm, font=, thick},
    edge/.style={-Stealth, thick, shorten >=2pt},
    msg/.style={-Stealth, dashed, thick, color=gray!70},
    agg/.style={rectangle, draw, fill=purple!10, minimum width=2.2cm, minimum height=0.9cm, font=, thick},
    edgefeat/.style={draw=none, fill=yellow!40, font=\scriptsize, inner sep=1pt},
    mlp/.style={rectangle, draw, fill=green!20, minimum width=1cm, minimum height=0.9cm, font=, thick},
    output/.style={circle, draw, fill=red!10, minimum size=0.5cm, font=, thick},
    plus/.style={circle, draw, fill=white, minimum size=0.5cm, font=, thick},
    annot/.style={font=\small, align=left}
]

\node[node] (i) at (0,0) {$x_i$};
\node[neighbor] (j1) at (-2,1.2) {$x_{j_1}$};
\node[neighbor] (j2) at (-2,-1.2) {$x_{j_2}$};
\node[neighbor] (j3) at (2,0) {$x_{j_3}$};

\draw[edge] (j1) -- node[edgefeat, midway, above left] {$e_{ij_1}$} (i);
\draw[edge] (j2) -- node[edgefeat, midway, below left] {$e_{ij_2}$} (i);
\draw[edge] (j3) -- node[edgefeat, midway, above right] {$e_{ij_3}$} (i);

\node[agg] (agg) at (2.2,-2) {$\sum\limits_{j \in \mathcal{N}(i)} \mathrm{ReLU}(x_j + e_{ij})$};
\draw[msg] (j1) .. controls (0.5,1.5) .. (agg.north);
\draw[msg] (j2) --  (agg.west);
\draw[msg] (j3) -- (agg.north);

\node[plus] (plus) at (0,-3.2) {$+$};

\draw[edge] (i) -- (plus);
\draw[edge] (agg) -- (plus);

\node[mlp] (mlp) at (0,-4.8) {MLP};
\draw[edge] (plus) -- (mlp);

\node[output, label={[annot]below:Updated embedding}] (out) at (0,-6.3) {$x'_i$};
\draw[edge] (mlp) -- (out);

\begin{scope}[shift={(2.2,-3.5)}, scale=0.5]
    \node[neighbor] (l1) at (0,0) {};
    \node[annot,right=0.2cm of l1] {Neighbor node};
    \node[node] (l2) at (0,-1.2) {};
    \node[annot,right=0.2cm of l2] {Central node};
    \node[output] (l3) at (0,-2.4) {};
    \node[annot,right=0.2cm of l3] {Output embedding};
    \draw[edge] (-0.4,-3.6) -- ++(1,0) node[annot,right=0.2cm] {Feature/aggregation flow};
    \draw[msg] (-0.4,-4.8) -- ++(1,0) node[annot,right=0.2cm] {\color{black} Message passing};
    \node[agg, minimum width=0.4cm, minimum height=0.4cm] (l4) at (0,-6) {};
    \node[annot,right=0.3cm of l4] {Aggregation};
    \node[plus, minimum size=0.4cm] (l5) at (0,-7.2) {+};
    \node[annot,right=0.2cm of l5] {Addition};
\end{scope}
\end{tikzpicture}
}
    }
    \caption{Message passing and aggregation of the GINEConv block.}
    \label{fig:gine_diagram}\vspace{-0.4cm}
\end{figure}

\subsection{\acf{gnn} as Feature Extractor}

\subsubsection{Node and Edge Feature Design}

The substrate network is modeled as a graph where nodes represent \acp{rh}, \acp{es}, and \acp{rc}, and edges denote physical links annotated with delay and bandwidth capacity. Each node is assigned features relevant to its functional role.

For non-\ac{rh} nodes (i.e., \acp{es} and \acp{rc}), features include their remaining resource capacity and a scalar \emph{node order}, which denotes the fixed index of the node in the graph. As \acp{gnn} are inherently permutation-invariant, they do not encode positional information about nodes by default. However, to enable the RL agent to take placement actions, the positional information is necessary. Therefore, the node order serves both as a positional inductive bias to help the GNN distinguish nodes and allow the RL agent to indicate the node index when taking placement actions.

For \acp{rh}, we do not include resource capacity, assuming that all \acp{rh} have sufficient internal capability to host the RU segment of any split. Instead, \ac{rh} are described by their service demands, including traffic load and end-to-end latency requirement. These demand attributes are also concatenated with their node order and processed separately.

Edge features consist of delay and bandwidth capacity. These are used during message passing to inform the model of topological constraints and link-specific characteristics. 

Once encoded, all node and edge features are projected to the same embedding dimension and passed to the \ac{gnn} layers for message passing.
Formally, let the substrate network be represented as a graph $G = (V, E)$, where $V$ is the set of nodes and $E$ is the set of edges. For each node $i \in V$, the node feature vector is defined as:
\begin{align}
    \mathbf{h}_i = [r_i, o_i, \lambda_i, \Delta_i]
\end{align}
where $r_i = R_i - p_i$ is the remaining resource capacity (for ES/RC nodes, $r_i = 0$ for RHs), $o_i$ is the node order (index), $\lambda_i$ is the traffic demand (for RHs, $\lambda_i = 0$ for ES/RCs), and $\Delta_i$ is the latency requirement (for RHs, $\Delta_i = 0$ for ES/RCs). The feature vector is constructed such that only relevant fields are non-zero for each node type.

For each edge $(i, j) \in E$, the edge feature vector is:
\begin{align}
    \mathbf{e}_{ij} = [b_{ij}, \delta_{ij}]
\end{align}
where $b_{ij}$ is the bandwidth capacity and $\delta_{ij}$ is the link delay between nodes $i$ and $j$. 
All node and edge features are projected to a common embedding dimension before being processed by the GNN layers.

\subsubsection{GNN Blocks}

Our feature extractor uses a two-layer Graph Isomorphism Network with Edge attributes (GINEConv) \cite{hu2020strategies} for the core blocks. This variant of GIN aggregates features from neighboring nodes with additive edge feature fusion. Particularly, for each node $i$, the GINEConv updates take the form:\vspace{-0.4cm}

\begin{equation}
    x'_i = \text{MLP}\left(x_i + \sum_{j \in \mathcal{N}(i)} \text{ReLU}(x_j + e_{ij})\right)
\end{equation}
where $x_i$ is the embedding of node $i$, $e_{ij}$ is the edge attribute between nodes $i$ and $j$, and $\mathcal{N}(i)$ is the set of neighbors. Two GINEConv layers are applied sequentially, with learnable MLPs and non-linear activations. This allows the model to propagate both node and edge information across the graph.

\subsubsection{Graph Embedding and Output Integration}
After message passing, the final node embeddings are aggregated using \emph{global mean pooling} to form a fixed-dimensional representation of the graph. This vector encodes the global state of the O-RAN substrate, including topology, resource availability, and service demands. 
The resulting graph-level embedding is provided to both the policy and value networks. This enables the RL agent to make globally informed joint decisions, i.e., assigning a functional split and placing the vDU and vCU for every \ac{rh} in a single forward pass, based on a unified, topology-aware state representation. 
By incorporating this GNN-based encoder, our architecture enables generalization across varying network sizes and topologies, and supports efficient learning in large, structured action spaces.

\section{Experiments}\label{sec:exp}

\subsection{Simulation set up}

\subsubsection{Overview}
We evaluate our approach on two O-RAN substrate scales: a small network (8 RHs, 3 ESs, 2 RCs) and a large network (64 RHs, 4 ESs, 2 RCs). We measure performance in terms of cumulative reward and total deployment cost, and test generalization on 5 different topologies with varied node distributions, connectivity patterns, and latency characteristics. All experiments share identical training conditions.

\subsubsection{Substrate Topology}
\begin{figure}
    \centering
    \subfloat[8 Base stations]{
        \includegraphics[width=0.45\linewidth]{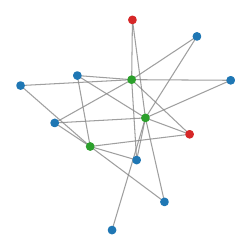}
    }
    \quad
    \subfloat[64 Base stations]{
        \includegraphics[width=0.45\linewidth]{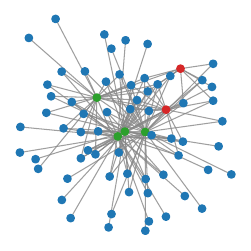}
    }
    \caption{Substrate network topologies (a) Small-scale network with 8 \acp{rh}, 3 \acp{es}, and 2 \acp{rc}. (b) Large-scale network with 64 \acp{rh}, 4 \acp{es}, and 2 \acp{rc}.}\vspace{-0.4cm}
    \label{fig:subnetworks}
\end{figure}

We generate two substrate topologies: (i) \textbf{Small-scale}: 8 RHs, 3 ESs, 2 RCs; (ii) \textbf{Large-scale}: 64 RHs, 4 ESs, 2 RCs. To ensure feasibility, each RH is connected to at least one ES and each ES to at least one RC, guaranteeing at least one valid RU--DU--CU path per RH. The generated topologies are shown in Figure~\ref{fig:subnetworks}.

\subsubsection{Split-4 Support}
For the legacy Split 4 configuration (omitting the DU), we add direct RH--RC links with a 10\% probability per RH. These links are provisioned to satisfy split-4 throughput and latency requirements.

\subsubsection{Resource Configuration}
Following prior work~\cite{murtiDeepReinforcementLearning2024, liNetMindAdaptiveRAN2024}, we assign ES and RC compute capacities of 20 and 100 \acp{cc}, respectively. Link bandwidths and latencies are sampled uniformly from [10, 40]~Gbps and [0, 3.6]~ms. Direct RH--RC links use 160~Gbps bandwidth and latencies in [0.1, 0.25]~ms to meet strict split-4 constraints.

\subsubsection{Request Generation}
At each timeslot, every RH issues a request consisting of traffic load and end-to-end latency requirement. We sample slice-specific SLA parameters from uniform distributions defined in~\cite{wuSlicingEnabledFlexible2024} and summarized in Table~\ref{tab:sim_req_stats}.

\subsubsection{Costing system}
The costing system is designed to reflect the deployment costs of the O-RAN system, including processing, routing, and reconfiguration costs. Each split $s$ has an associated processing cost $\psi^{\rm DU}_s$ for vDU deployment on ESs and $\psi^{\rm CU}_s$ for vCU deployment on RCs. For split $s \in \{1, 2, 3, 4\}$, each vDU incurs a processing cost of $\psi^{\rm DU}_s = \{0.05, 0.04, 0.00325, 0\}$ \acp{cc} per Mbps, and $\psi^{\rm CU}_s = \{0, 0.001, 0.00175, 0.05\}$ \acp{cc} per Mbps for each vCU, respectively, as in \cite{murtiConstrainedDeepReinforcement2022, amiriDeepReinforcementLearning2024}. Routing and reconfiguration costs are defined as in Equations \ref{eq:reconfig_cost} and \ref{eq:routing_cost}. In experiments, we set $\phi_{\rm R} = 1$ and $\phi_{\rm L} = 1$, but in reality, these can be adjusted to reflect different cost structures.
\begin{table}[]
    \centering
    \caption{Simulation request statistics for different network slices.}
    \label{tab:sim_req_stats}
    \begin{tabular}{@{}lcc@{}}
        \toprule
        \textbf{Network slice}  & \textbf{Traffic load (Mbps)} & \textbf{End-to-end latency (ms)}  \\ \midrule
        eMBB & U(250, 300) & U(15, 20) \\
        mMTC & U(150, 200) & U(180, 200) \\
        uRLLC & U(20, 40) & U(2, 4) \\ \bottomrule
    \end{tabular}\vspace{-0.4cm}
\end{table}

\subsection{Evaluation setup}

\subsubsection{Baseline Methods}
We conduct comprehensive comparisons against three state-of-the-art baseline methods to evaluate the effectiveness of our proposed GPPO approach:
\begin{itemize}
  \item \textbf{DDPG}~\cite{lillicrap2015continuous}: A deep deterministic policy gradient algorithm implemented with an MLP-based actor-critic architecture. The continuous action outputs are rescaled and rounded to transform into discrete actions.
  \item \textbf{PPO}~\cite{schulman2017ppo}: The standard proximal policy optimization algorithm, using an MLP-based policy network that directly operates on the discrete action space without action masking.
  \item \textbf{MPPO}~\cite{huang2020closer,sb3-contrib}: PPO with invalid action masking to improve learning efficiency by preventing selection of infeasible actions.
  \item \textbf{GPPO} (ours): Our proposed method that augments MPPO with a GNN-based feature extractor to enable topology-aware state representation and decision-making.
\end{itemize}

\subsubsection{Implementation Details}
All methods are implemented in Python using the Stable-Baselines3~\cite{stable-baselines3} framework and its contrib package~\cite{sb3-contrib}. We employ Optuna~\cite{optuna_2019} for automated hyperparameter optimization across all baseline methods to ensure fair comparison. The training infrastructure utilizes 32 parallel environments executed on 12 CPU cores with SubprocVecEnv vectorization for efficient sample collection.

\subsubsection{Training Configuration}
Each agent is trained for 600\,000 timesteps across both network scales. Training episodes span 288 time slots each, with a resource releasing ratio of 0.5 to maintain dynamic demand patterns. The environment generates requests following the slice-specific distributions detailed in Table~\ref{tab:sim_req_stats}.

\subsubsection{Hyperparameter Settings}
Through Optuna optimization, we identified the following key hyperparameters for the PPO-based methods (PPO, MPPO, and GPPO): learning rate $\alpha = 10^{-4}$, batch size of 128 samples, discount factor $\gamma = 0.98$, GAE lambda $\lambda = 0.97$, clipping range $\epsilon = 0.3$, and entropy coefficient $c_{\text{ent}} = 10^{-6}$. DDPG employs independently optimized hyperparameters following standard actor-critic configurations with experience replay and target networks.

\subsubsection{Network Architectures}
The MLP-based policy and value networks in MPPO and GPPO consist of two hidden layers with 256 units each, using ReLU activation functions. Our GPPO method additionally incorporates a GNN feature extractor with 1024 hidden features, implemented using two-layer GINEConv blocks as described in Section~4.3. All networks are initialized using Xavier uniform initialization~\cite{glorot2010understanding}.

\subsubsection{Evaluation Protocol}
In the normal setup, we evaluate trained models on the same static topologies used during training to assess learning effectiveness. Each model performs 10 evaluation episodes per network scale, and we report the mean and standard deviation of cumulative reward and total deployment cost. For the large-scale scenarios, we include only episodes achieving feasible deployments (no connectivity constraint violations) in the cost analysis to ensure fair comparison. In the generalization setup, we trained a \ac{drl} model on 5 different 8-\ac{rh} topologies simultaneously and evaluated them on the same 5 topologies for 100 evaluation episodes in total. All reported results represent averages over six independent training runs with different random seeds to ensure statistical reliability and account for training variance.

\subsection{Results}

We evaluate performance primarily using two metrics: reward and deployment cost. Mean rewards include all episodes regardless of feasibility, as negative rewards from constraint violations provide important information about method reliability. However, deployment costs are only reported for episodes that achieve feasible solutions, since infeasible deployments cannot be physically implemented and thus have no meaningful cost.

\subsubsection{Small scale}
\begin{figure}
    \centering
    \subfloat[Deployment cost]{
        \begin{tikzpicture}
\begin{axis}[
    width=128pt,
    height=110pt,
    xbar interval=12pt,
    grid=major,
    grid style={draw=gray!20},
    bar width=8pt,
    symbolic y coords={PPO, DDPG, MPPO, GPPO},
    ytick=data,
    xmin=0, xmax=200,
    yticklabels={},
    x tick label style={font=\footnotesize},
    axis x line*=bottom,
    axis y line=none,
    tick label style={font=\footnotesize},
    title style={font=\footnotesize},
    clip=false,
]

\addplot[
    xbar,
    fill=red!70,
    draw=none,
    line width=0pt,
    error bars/.cd,
    x dir=both,
    x explicit,
    error bar style={line width=0.8pt},
    error mark options = {
        rotate = 90,
        line width=0.8pt, 
        mark size = 2pt,
    }
] coordinates {(89.06,PPO) +- (4.7632859188,4.7632859188)};

\addplot[
    xbar,
    fill=orange!70,
    draw=none,
    error bars/.cd,
    x dir=both,
    x explicit,
    error bar style={line width=0.8pt},
    error mark options = {
        rotate = 90,
        line width=0.8pt, 
        mark size = 2pt,
    }
] coordinates {(193.76,DDPG) +- (1.8671643813,1.8671643813)};

\addplot[
    xbar,
    fill=cyan!80!blue,
    draw=none,
    error bars/.cd,
    x dir=both,
    x explicit,
    error bar style={line width=0.8pt},
    error mark options = {
        rotate = 90,
        line width=0.8pt, 
        mark size = 2pt,
    }
] coordinates {(84.94,MPPO) +- (3.2468843862,3.2468843862)};

\addplot[
    xbar,
    fill=violet!40!pink,
    draw=none,
    error bars/.cd,
    x dir=both,
    x explicit,
    error bar style={line width=0.8pt},
    error mark options = {
        rotate = 90,
        line width=0.8pt, 
        mark size = 2pt,
    }
] coordinates {(78.23,GPPO) +- (3.5012230712,3.5012230712)};

\node[anchor=south west,font=\footnotesize,yshift=2pt,align=left] at (axis cs:0,PPO) {PPO};
\node[anchor=south west,font=\footnotesize,yshift=2pt,align=left] at (axis cs:0,DDPG) {DDPG};
\node[anchor=south west,font=\footnotesize,yshift=2pt,align=left] at (axis cs:0,MPPO) {MPPO};
\node[anchor=south west,font=\footnotesize,yshift=2pt,align=left] at (axis cs:0,GPPO) {GPPO};

\end{axis}
\end{tikzpicture}
    }
    \quad
    \subfloat[Training reward]{
        \begin{tikzpicture}
\begin{axis}[
    width=128pt,
    height=110pt,
    xbar interval=12pt,
    grid=major,
    grid style={draw=gray!20},
    bar width=8pt,
    symbolic y coords={PPO, DDPG, MPPO, GPPO},
    ytick=data,
    xmin=0, xmax=90,
    xtick={0, 20, 40, 60, 80},
    yticklabels={},
    x tick label style={font=\footnotesize},
    axis x line*=bottom,
    axis y line=none,
    tick label style={font=\footnotesize},
    title style={font=\footnotesize},
    clip=false,
]

\addplot[
    xbar,
    fill=red!70,
    draw=none,
    line width=0pt,
    error bars/.cd,
    x dir=both,
    x explicit,
    error bar style={line width=0.8pt},
    error mark options = {
        rotate = 90,
        line width=0.8pt, 
        mark size = 2pt,
    }
] coordinates {(84.88,PPO) +- (1.3410,1.3410)};

\addplot[
    xbar,
    fill=orange!70,
    draw=none,
    error bars/.cd,
    x dir=both,
    x explicit,
    error bar style={line width=0.8pt},
    error mark options = {
        rotate = 90,
        line width=0.8pt, 
        mark size = 2pt,
    }
] coordinates {(68.2572,DDPG) +- (0.1518,0.1518)};

\addplot[
    xbar,
    fill=cyan!80!blue,
    draw=none,
    error bars/.cd,
    x dir=both,
    x explicit,
    error bar style={line width=0.8pt},
    error mark options = {
        rotate = 90,
        line width=0.8pt, 
        mark size = 2pt,
    }
] coordinates {(86.9035,MPPO) +- (1.1813,1.1813)};

\addplot[
    xbar,
    fill=violet!40!pink,
    draw=none,
    error bars/.cd,
    x dir=both,
    x explicit,
    error bar style={line width=0.8pt},
    error mark options = {
        rotate = 90,
        line width=0.8pt, 
        mark size = 2pt,
    }
] coordinates {(88.9189,GPPO) +- (1.0178,1.0178)};

\node[anchor=south west,font=\footnotesize,yshift=2pt,align=left] at (axis cs:0,PPO) {PPO};
\node[anchor=south west,font=\footnotesize,yshift=2pt,align=left] at (axis cs:0,DDPG) {DDPG};
\node[anchor=south west,font=\footnotesize,yshift=2pt,align=left] at (axis cs:0,MPPO) {MPPO};
\node[anchor=south west,font=\footnotesize,yshift=2pt,align=left] at (axis cs:0,GPPO) {GPPO};

\end{axis}
\end{tikzpicture}
    }
    \caption{Performance comparison on the 8-\ac{rh} topology: (a) deployment cost and (b) training reward.}\vspace{-0.4cm}
    \label{fig:benchmark_8}
\end{figure}
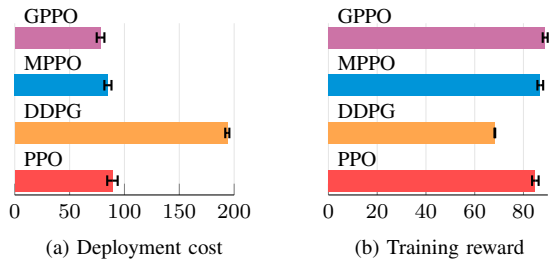

We first evaluate our approach on the small-scale topology with 8 \acp{rh}. \Figref{fig:benchmark_8} presents the comparative performance across all baseline methods.
The results demonstrate clear performance differences across methods. As detailed in Table~\ref{tab:results_summary}, GPPO consistently outperforms all baseline methods across both performance metrics and network scales. 
For the small-scale topology, GPPO achieves the best overall performance in both reward and deployment cost metrics. MPPO follows closely in reward performance but incurs slightly higher deployment costs. PPO shows moderate performance across both metrics, while DDPG significantly underperforms with the highest costs and lowest rewards. The performance gap between GPPO and MPPO on this simple topology is modest, with GPPO achieving a \textbf{7.9\%} cost reduction while maintaining similar reward levels. 
As shown in Figure~\ref{fig:ep_re}, GPPO shows slightly faster convergence on the 8-\ac{rh} topology compared to MPPO, though both methods achieve comparable final performance.\vspace{-0.3cm}

\begin{table}[h]
\centering
\begin{threeparttable}
\caption{Quantitative performance comparison across network scales and methods.}
\label{tab:results_summary}
\setlength{\tabcolsep}{3pt}
\begin{tabular}{lcccc}
\toprule
\textbf{Topology} & \textbf{Method} & \textbf{Mean Reward} & \textbf{Deploy Cost} & \makecell{\textbf{Success}\\\textbf{Rate}} \\
\midrule
\multirow{4}{*}{\makecell{Small-scale\\(8 \acp{rh})}} 
& DDPG & 68.26 {\scriptsize $\pm$ 0.15} & 193.76 {\scriptsize $\pm$ 1.87} & 6/6 \\
& PPO & 84.88 {\scriptsize $\pm$ 1.34} & 89.06 {\scriptsize $\pm$ 4.76} & 6/6 \\
& MPPO & 86.90 {\scriptsize $\pm$ 1.18} & 84.94 {\scriptsize $\pm$ 3.25} & 6/6 \\
& \textbf{GPPO (ours)} & \textbf{88.92} {\scriptsize $\pm$ 1.02} & \textbf{78.23} {\scriptsize $\pm$ 3.50} & \textbf{6/6} \\
\midrule
\multirow{4}{*}{\makecell{Large-scale\\(64 \acp{rh})}} 
& DDPG* & $-$1.57 {\scriptsize $\pm$ 0.05} & --- & 0/6 \\
& PPO* & $-$1.13 {\scriptsize $\pm$ 0.03} & --- & 0/6 \\
& MPPO & 24.28 {\scriptsize $\pm$ 11.34} & 7,982.97 {\scriptsize $\pm$ 531.18} & 3/6 \\
& \textbf{GPPO (ours)} & \textbf{52.48} {\scriptsize $\pm$ 0.36} & \textbf{6,528.84} {\scriptsize $\pm$ 166.38} & \textbf{6/6} \\
\bottomrule
\end{tabular}
\begin{tablenotes}
\item *: Deployment costs unavailable for infeasible solutions since no actual deployment occurs.
\end{tablenotes}
\end{threeparttable}
\end{table}

\subsubsection{Large scale}
\begin{figure}
    \centering
    \subfloat[Deployment cost]{
        \begin{tikzpicture}
\begin{axis}[
    width=128pt,
    height=110pt,
    enlarge y limits=0.6,
    xbar interval=12pt,
    grid=major,
    grid style={draw=gray!20},
    clip=false,
    bar width=10pt,
    symbolic y coords={MPPO, GPPO},
    ytick=data,
    xmin=0, xmax=8800,
    yticklabels={},
    scaled ticks=base 10:-3,
    x tick label style={font=\small},
    axis x line*=bottom,
    axis y line=none,
    tick label style={font=\small},
    title style={font=\footnotesize},
    every x tick label/.append style={alias=XTick,inner xsep=0pt},
    every x tick scale label/.style={at=(XTick.base east),anchor=base west}
]

\addplot[
    xbar,
    fill=cyan!80!blue,
    draw=none,
    line width=0pt,
    error bars/.cd,
    x dir=both,
    x explicit,
    error bar style={line width=0.8pt},
    error mark options = {
        rotate = 90,
        line width=0.8pt, 
        mark size = 3pt,
    }
] coordinates {(7982.97,MPPO) +- (531.18,531.18)};

\addplot[
    xbar,
    fill=violet!40!pink,
    draw=none,
    error bars/.cd,
    x dir=both,
    x explicit,
    error bar style={line width=0.8pt},
    error mark options = {
        rotate = 90,
        line width=0.8pt, 
        mark size = 3pt,
    }
] coordinates {(6528.84,GPPO) +- (166.38,166.38)};

\node[anchor=south west,font=\footnotesize,yshift=3pt] at (axis cs:0,MPPO) {MPPO};
\node[anchor=south west,font=\footnotesize,yshift=3pt] at (axis cs:0,GPPO) {GPPO};

\end{axis}
\end{tikzpicture}
    }
    \subfloat[Training reward]{
        \begin{tikzpicture}
\begin{axis}[
    width=150pt,
    height=110pt,
    xbar interval=12pt,
    grid=major,
    grid style={draw=gray!20},
    bar width=8pt,
    symbolic y coords={PPO, DDPG, MPPO, GPPO},
    ytick=data,
    xmin=-5, xmax=58,
    xtick={0, 10, 20, 30, 40, 50},
    yticklabels={},
    x tick label style={font=\footnotesize},
    axis x line*=bottom,
    axis y line=none,
    tick label style={font=\footnotesize},
    title style={font=\footnotesize},
    clip=false,
]

\addplot[
    xbar,
    fill=red!70,
    draw=none,
    line width=0pt,
    error bars/.cd,
    x dir=both,
    x explicit,
    error bar style={line width=0.8pt},
    error mark options = {
        rotate = 90,
        line width=0.8pt, 
        mark size = 2pt,
    }
] coordinates {(-1.13,PPO) +- (0.030297999109,0.030297999109)};

\addplot[
    xbar,
    fill=orange!70,
    draw=none,
    error bars/.cd,
    x dir=both,
    x explicit,
    error bar style={line width=0.8pt},
    error mark options = {
        rotate = 90,
        line width=0.8pt, 
        mark size = 2pt,
    }
] coordinates {(-1.57,DDPG) +- (0.0539,0.0539)};

\addplot[
    xbar,
    fill=cyan!80!blue,
    draw=none,
    error bars/.cd,
    x dir=both,
    x explicit,
    error bar style={line width=0.8pt},
    error mark options = {
        rotate = 90,
        line width=0.8pt, 
        mark size = 2pt,
    }
] coordinates {(24.28,MPPO) +- (11.33899125,11.33899125)};

\addplot[
    xbar,
    fill=violet!40!pink,
    draw=none,
    error bars/.cd,
    x dir=both,
    x explicit,
    error bar style={line width=0.8pt},
    error mark options = {
        rotate = 90,
        line width=0.8pt, 
        mark size = 2pt,
    }
] coordinates {(52.48,GPPO) +- (0.35643263633,0.35643263633)};

\node[anchor=south west,font=\footnotesize,yshift=2pt,align=left] at (axis cs:-5,PPO) {PPO};
\node[anchor=south west,font=\footnotesize,yshift=2pt,align=left] at (axis cs:-5,DDPG) {DDPG};
\node[anchor=south west,font=\footnotesize,yshift=2pt,align=left] at (axis cs:-5,MPPO) {MPPO};
\node[anchor=south west,font=\footnotesize,yshift=2pt,align=left] at (axis cs:-5,GPPO) {GPPO};

\end{axis}
\end{tikzpicture}
    }
    \caption{Performance comparison on the 64-\ac{rh} topology: (a) deployment cost for successful deployments and (b) training reward.}
    \label{fig:benchmark_64}\vspace{-0.4cm}
\end{figure}
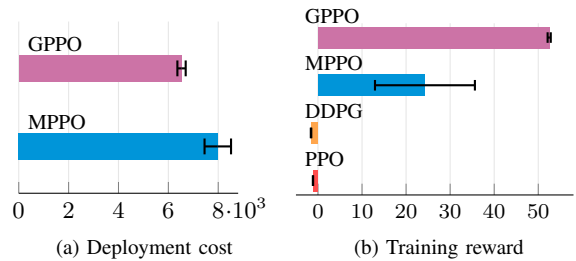

To assess scalability, we evaluate our method on the large-scale topology with 64 \acp{rh}. \Figref{fig:benchmark_64} demonstrates the performance under increased network complexity.

The large-scale results reveal significant scalability challenges for traditional methods. Most critically, PPO and DDPG fail to achieve consistent feasible deployments, with all training runs converging to negative rewards, indicating complete failure to satisfy connectivity constraints. For these failed runs, the negative rewards are included in the reported mean performance, but no deployment costs can be computed since no actual network deployment occurs when constraints are violated. MPPO shows limited scalability, with only 3 out of 6 runs achieving feasible solutions. Moreover, even when MPPO does produce feasible deployment policies, these solutions are substantially suboptimal in cost efficiency.

In stark contrast, our proposed GPPO achieves a perfect success rate across all training runs with substantially higher mean rewards than MPPO. This dramatic performance gap highlights the critical importance of graph-based feature extraction for complex network topologies. Notably, GPPO not only ensures consistent feasibility but also delivers superior solution quality, achieving an \textbf{18.2\%} cost reduction while maintaining perfect reliability, as detailed in Table~\ref{tab:results_summary}.

\begin{figure}
    \centering
    \includegraphics[width=0.6\linewidth]{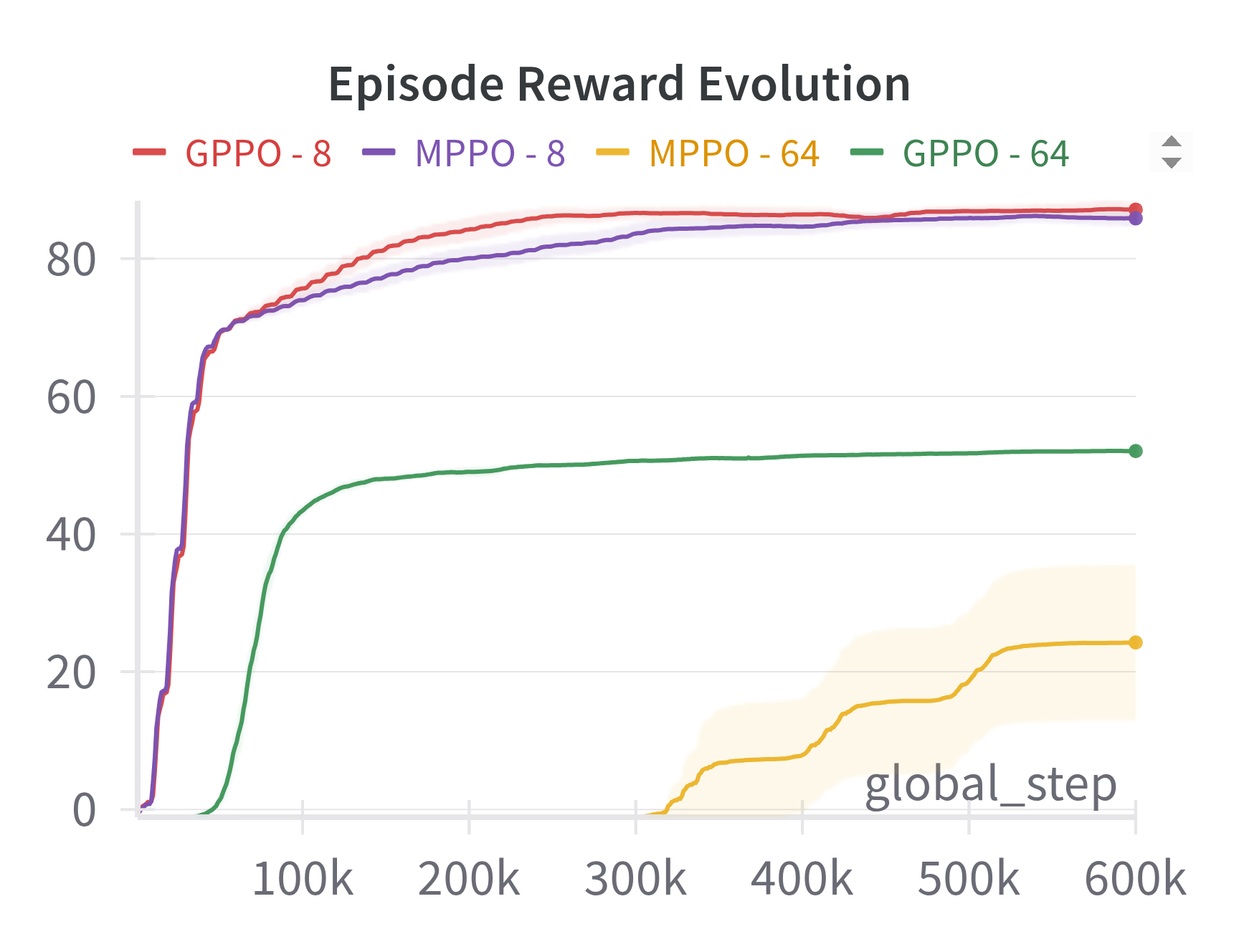}\vspace{-0.4cm}
    \caption{Reward evolution during training across methods and network scales.}\vspace{-0.4cm}
    \label{fig:ep_re}
\end{figure}

Figure~\ref{fig:ep_re} reveals that PPO and DDPG fail completely on the 64-\ac{rh} topology, while MPPO shows delayed learning after 350k steps with high variance. GPPO maintains consistent learning from early stages, demonstrating the GNN's effectiveness in extracting topological patterns for complex scenarios.

\subsubsection{Generalization}
\begin{table}[h]
\centering
\caption{Generalization performance comparison on different O-RAN topologies.}
\label{tab:generalization_results}
\setlength{\tabcolsep}{6pt}
\begin{tabular}{lcc}
\toprule
\textbf{Method} & \textbf{Mean Cost} & \textbf{Mean Reward} \\
\midrule
MPPO & $261.46 \pm 92.50$ & $48.47 \pm 12.57$ \\
GPPO (ours) & $211.80 \pm 63.09$ & $60.57 \pm 10.72$ \\
\bottomrule
\end{tabular}
\end{table}
As expected, the generalization performance of both models has significant variance, due to the diverse topological characteristics of the 5 topologies. However, GPPO consistently outperforms MPPO across all metrics, achieving a \textbf{19\% }cost reduction and \textbf{25\%} reward improvement on average, as shown in Table~\ref{tab:generalization_results}. This demonstrates GPPO's superior ability to generalize learned policies to unseen network structures, validating the effectiveness of graph-based feature extraction for topology-aware decision-making.

\subsubsection{Key Findings}
Our experimental evaluation reveals several important insights:

\textbf{Generalization Capability:} GPPO demonstrates a clear advantage in generalizing to unseen O-RAN topologies, consistently outperforming MPPO in both cost and reward metrics. This highlights the effectiveness of graph-based feature extraction for robust, topology-aware decision-making in dynamic network environments.

\textbf{Scalability Advantage:} The most significant finding is GPPO's superior scalability. While multiple methods achieve reasonable performance on the small-scale topology, only graph-enhanced approaches maintain effectiveness as network complexity increases. The complete failure of PPO and DDPG on the large-scale topology, combined with MPPO's poor reliability, demonstrates the fundamental limitations of traditional \ac{rl} methods for complex O-RAN optimization.

\textbf{Reliability vs Performance Trade-off:} Beyond mean performance metrics, success rates reveal critical differences in method reliability. GPPO achieves perfect reliability across all scales, while MPPO's 50\% success rate on large topologies makes it unsuitable for production deployment despite reasonable performance when successful.

\textbf{Cost-Performance Synergy:} Notably, GPPO achieves both higher rewards and lower costs across all scenarios, as shown in Table~\ref{tab:results_summary}. On the small-scale topology, GPPO's 2.3\% reward improvement over MPPO comes with 7.9\% cost reduction. At scale, this advantage reduces to 18.2\% lower costs. This demonstrates that graph-enhanced learning not only ensures reliable feasibility but also leads to fundamentally superior optimization solutions, outperforming traditional methods even in their successful deployment cases.

\textbf{Action Masking Benefits:} The consistent improvement of MPPO over vanilla PPO validates the importance of constraint-aware action selection, though this benefit diminishes at scale without topological awareness.

\section{Conclusion} \label{sec:conclusion}
This paper presented a scalable framework for O-RAN resource management based on Graph-Augmented Proximal Policy Optimization. By jointly optimizing functional split selection and virtualized baseband unit placement, and leveraging \ac{gnn}-based feature extraction with action masking, our approach effectively addresses the complexity and dynamics of real-world O-RAN environments. Experimental results on two different network scales demonstrate that GPPO achieves superior performance, reliability, and generalization compared to existing methods, reducing deployment costs and improving reward metrics even when handling different topologies at the same time. These findings highlight the importance of topology-aware learning and joint optimization for future O-RAN systems. Future work will explore further improvements in scalability, adaptation to online network changes, and integration with end-to-end network orchestration frameworks.

\bibliographystyle{IEEEtran}
\bibliography{citation}
\end{document}